\documentclass[12pt]{article}
\usepackage{amsmath}
\usepackage{amssymb}
\usepackage{amsthm}
\usepackage{amsfonts}
\usepackage{graphicx}
\usepackage{textcomp}
\usepackage{color}
\usepackage{hyperref}
\usepackage{tikz-cd}
\numberwithin{equation}{section}

\begin{document}

\begin{titlepage}

\title{Weyl Covariant Theories of Gravity in Riemann-Cartan-Weyl Space-times \\ I. Topologically Massive Gravity}

\author{ Tekin Dereli\footnote{tdereli@ku.edu.tr}, Cem Yeti\c{s}mi\c{s}o\u{g}lu\footnote{cyetismisoglu@ku.edu.tr} \\ {\small Department of Physics, Ko\c{c} University, 34450 Sar{\i}yer, \.{I}stanbul, Turkey }}

\date{17 April 2019}

\maketitle

\begin{abstract}

\noindent We discuss locally Weyl (scale) covariant generalisations of gravitational theories using Riemann-Cartan-Weyl space-times in arbitrary dimensions. We demostrate the procedure of Weyl gauging on two examples in particular: General relativity, and Topologically Massive Gravity in three dimensions.
\end{abstract}

\vskip 2cm

\noindent {\bf Keywords}: Riemann-Cartan-Weyl Spaces $\cdot$ Scale Invariance $\cdot$ Massive Gravity Theories

\thispagestyle{empty}

\end{titlepage}

\maketitle			
\clearpage 

\section{Introduction}
\setcounter{page}{1}

\noindent General Relativity (GR) needs to be modified in both large and small scales. On one hand, probable deviations from geodesic motion due to gravitational self forces and several astrophysical observations hint modifications at large scales. On the other hand, a renormalizable, unitary quantum field theory of massless spin-2 gravitons insists on a modification at short distances. Scale covariant generalisations of gravitational theories are important to consider both UV and IR complete theories of gravitation. Here, we give a procedure for Weyl gauging of gravitational theories in the context of Riemann-Cartan-Weyl (RCW) space-times. We show that our method gives a consistent generalisation for two examples: GR theory in arbitrary dimensions and Topologically Massive Gravity (TMG) theory in three dimensions. We picked a three dimensional example because three dimensional theories of gravitation provide important toy models to understand quantum gravity problem [1-4]. TMG theory is an extension of GR in three dimensions that is obtained by coupling the gravitational Chern-Simons term to the Einstein-Hilbert action [5, 6]. Because standart general relativity has no local degree of freedom [7], Chern-Simons term introduces local degrees of freedom by increasing the order of the field equations from two to three. TMG admits a massive spin-2 mode in addition to a famous BTZ blackhole solution [8, 9]. Furthermore, using AdS/CFT correspondence, unitarity of the renormalisable [10] quantum theory has been shown at a critical point [11].  \\

\noindent RCW space-times are geometries whose linear connections have both dynamic torsion and non-metricity parts. The particular choice of connection where the non-metricity tensor has a vanishing trace-free part provides a natural framework to discuss locally scale covariant theories of gravitation [12]. This is because the non-metricity tensor in a RCW space-time can be identified with the Weyl connection 1-form. This identification also relates the origin of local changes of scale to the geometry of space-time. \\

\noindent Weyl group is the group of local scale transformations. It is a non-compact, 1-parameter, abelian Lie group. As a set, it is homeomorphic to $\mathbb{R}^+$. For scale covariant theories, we promote global scale transformations into local ones. Therefore when we talk about scale covariant field theories, there is a principal bundle structure over space-time where the structure group is Weyl group. Tensors over space-time are seen as sections of this bundle. Transformations of such sections under this group is connected to the dimensions that they carry. On this bundle, a Weyl connection 1-form taking values in the Lie algebra $\mathbb{R}$ of the Weyl group introduces a scale covariant exterior derivative that is compatible with the action of the Weyl group. \\

\noindent The organisation of the paper is as follows. In section 2, we discuss RCW space-times within the context of local scale transformations. In section 3 we move on to give our examples. We present the Lagrangian formulation of locally scale covariant extension of these theories. Their field equations are determined using a first order variational principle. Finally, we show the consistency of this procedure and provide concluding remarks.
 
\section{Riemann-Cartan-Weyl Space-Times}

\noindent A RCW space-time is a triplet $\{M, g, \nabla\}$ where $M$ denotes a smooth $n$-manifold, $g$  a non-degenerate metric tensor, and  $\nabla$ a linear connection on $M$. The metric tensor can be expressed as $g=\eta_{ab} e^a \otimes e^b$ where $\eta_{ab} = g(X_a,X_b)= (-,+,..,+)$, in terms of a $g$-orthonormal frame $\{X_a\}$ that are dual to the co-frame 1-forms $\{e^a\}$ so that $e^a(X_b)=\iota_b e^a = \delta^a_b$. Here $\iota_a\equiv\iota_{X_a}$ denotes the interior product operations with respect to the frame vectors $X_a$. The space-time orientation is fixed by the choice of a volume form $*1= e^0 \wedge e^1 \wedge ... \wedge e^{n-1}$ where $*:\Lambda^p(M) \to \Lambda^{n-p}(M)$ is the Hodge duality operator. For simplicity, the following abbreviations for the exterior products $e^{ab\dots}\equiv e^a \wedge e^b \wedge \dots $ , and the interior products $\iota_{ab\dots} \equiv \iota_a \iota_b \dots$ will be used. Lastly, the connection $\nabla$ is given by a set of connection 1-forms $\{\Lambda^a_{\ b}\}$ so that $\nabla_{X_a} X_b = \Lambda^c_{\ b} (X_a) X_c$. We define the non-metricity, torsion, and curvature forms of a linear connection by the Cartan's structure equations below:
\begin{align}
\overset{(\Lambda)}{D}\eta_{ab}&= -(\Lambda_{ab} + \Lambda_{ba}) = -2Q_{ab},\\
\overset{(\Lambda)}{D}e^a &= de^a + \Lambda^a_{\ b} \wedge e^b = T^a, \\
\overset{(\Lambda)}{D}\Lambda^a_{\ b} &=d\Lambda^a_{\ b} +\Lambda^a_{\ c} \wedge\Lambda^c_{\ b} = \overset{(\Lambda)}{R^a_{\ b}}.
\end{align}
$d$, $\overset{(\Lambda)}{D}$ and $\overset{(\Lambda)}{R^a_{\ b}}$ denote the exterior derivative, exterior covariant derivative and curvature of the above connection, respectively.\footnote{Since connection is not metric-compatible, in RCW space-times the index lowering and raising operations do not commute with the exterior covariant derivative in general.} Bianchi identities are obtained as the integrability conditions of the above Cartan's structure equations:
\begin{align}
\overset{(\Lambda)}{D}Q_{ab}&= \frac{1}{2} (\overset{(\Lambda)}{R_{ab}} + \overset{(\Lambda)}{R_{ba}}), \\
\overset{(\Lambda)}{D}T^a &= \overset{(\Lambda)}{R^a_{\ b}} \wedge e^b,\\
\overset{(\Lambda)}{D}\overset{(\Lambda)}{R^a_{\ b}} &=0.
\end{align}

\noindent The most general linear connection is fixed uniquely by the metric tensor field $g$, the torsion tensor field $T$ and a non-metricity tensor field $S=\overset{(\Lambda)}{D} g$. To observe this, one may separate the anti-symmetric and symmetric parts of the connection 1-forms as follows: 
\begin{equation}
\Lambda^a_{\ b} = \Omega^a_{\ b} + Q^a_{\ b},
\end{equation}
where the anti-symmetric part further decomposes in a unique way according to
\begin{equation}
\Omega^a_{\ b} = \omega^a_{\ b} + K^a_{\ b} + q^a_{\ b}.
\end{equation}
Here,  the Levi-Civita connection 1-forms $\omega^a_{\ b}$ are determined completely by the co-frames from the Cartan structure equations
\begin{equation}
de^a+  \omega^a_{\ b} \wedge e^b = 0.
\end{equation}
The contortion 1-forms $K^a_{\ b}$ are fixed by the torsion 2-forms
\begin{equation}
K^a_{\ b}\wedge e^b = T^a .
\end{equation}
The anti-symmetric 1-forms $q^a_{\ b}$ are completely determined in terms of the symmetric non-metricity 1-forms $Q^a_{\ b}$ by the equations
\begin{equation}
q^a_{\ b}=-(\iota^a Q_{bc}) e^c + (\iota_b Q^a_{\ c}) e^c .
\end{equation}

\medskip

\noindent RCW geometry offers a framework to develop locally scale covariant generalisation of gravitational theories. Some field elements $\Phi$ are allowed to carry some representation of this group. Under a local scale transformation, we assume
\begin{equation}
\Phi \mapsto \exp(-q\sigma) \Phi,
\end{equation}
where $\sigma$ is a dimensionless real scalar field on space-time and the dimensionless parameter $q$ is called the Weyl charge. Weyl charges are assigned to fields according to the dimensions they are carrying [12, 13]. Conventionally, the metric tensor is assigned a Weyl charge of $-2$ because it is a covariant 2-tensor. This also complements the fact that a metric has dimension length squared. Then, Weyl charge assignments of other tensorial quantities are determined according to this choice. Therefore, Weyl charges are related to the equivalence class of metric tensors $[g]$ that carry a representation of the Weyl group.\footnote{This equivalence class defines a conformal structure on space-time. Two conformally equivalent metrics have the same distribution of light-cones.} The representatives of $[g]$ are equivalent under
\begin{equation}
g \mapsto \exp(2\sigma) g \label{meteq}.
\end{equation}
For the linear connection, we adopt the Weyl transformation rule under Weyl group action, that is
\begin{equation}
\nabla \mapsto \nabla.
\end{equation}
This choice is consistent with our framework because connection is not a tensorial quantity and thus it is not assigned any dimensions, so it stays inert under local scale transformations. Also, there need not be any correlations between metric scaling and transformation of the linear connection in a RCW space. To be able to make this choice, one has to have at least one of the torsion or non-metricity tensors to be nonzero. Otherwise, in a Riemannian space-time, the transformation property of the connection is determined by the metric tensor only. \\

\noindent Space-time exterior covariant derivative does not transform covariantly under local changes of scale. Hence, a Weyl connection is introduced as a compensating potential 1-form. A Weyl connection is a dimensionless 1-form $Q$\footnote{Here we denote the Weyl connection with $Q$ because later it will be identified with the trace of connection 1-forms $\Lambda^a_{\ b}$.} which transforms under a local scale transformation as
\begin{equation}
Q \mapsto Q + d\sigma .
\end{equation}
With the help of $Q$, the exterior Weyl covariant derivative of a p-form $\Phi^p_q$ with Weyl charge $q$ is defined by:
\begin{equation}
\mathcal{D} \Phi^p_q = \overset{(\Lambda)}{D} \Phi^p_q + qQ \wedge \Phi^p_q    \label{wcd}
\end{equation}
so that under a local scale transformation $\mathcal{D} \Phi^p_q \mapsto \exp(-q\sigma) \mathcal{D} \Phi^p_q$. \\

\noindent For every class $[g]$ of metric tensors, a class $[*]$ of Hodge maps is associated. Then, under the action of Hodge map, one has
\begin{align}
* \Phi^p_q &= \Phi^{n-p}_{q - (n-2p)}, \\
\mathcal{D} * \Phi^p_q &= \overset{(\Lambda)}{D} *\Phi^p_q + (q - (n-2p))Q \wedge * \Phi^p_q.
\end{align}
In addition, under contractions with the interior product $\iota_a$, Weyl charge of the fields increase by 1, i.e.
\begin{equation}
\iota_a \Phi^p_q = \Phi^{p-1}_{q+1}.
\end{equation}
In order to construct a locally scale invariant action, we take the following relation between the dynamic Weyl connection 1-form $Q$ and the non-metricity tensor $S=\overset{(\Lambda)}{D} g$:
\begin{equation}
S = 2Q \wedge g.
\end{equation}
This choice can be equivalently expressed as $\mathcal{D}g=\overset{(\Lambda)}{D}g-2Q\wedge g =0$.\footnote{In a Riemannian space-time, this equation is a compatibility condition between the conformal and projective structures. The projective structure is an equivalence relation between connections $\omega^a_{\ b}$ and $\omega^a_{\ b}+ \psi^a e_b - \psi_b e^a$ where $\psi=\psi^ae_a$ is a 1-form. Two projectively equivalent connections have the same geodesics up to reparametrization. For details, we refer to [14].} Thus, in a geometry equipped with the above features, the Weyl connection 1-form $Q$ and the non-metricity 1-forms $Q_{ab}$ are related by
\begin{equation}
Q_{ab} = -Q \eta_{ab} \label{nonmet}.
\end{equation}
This identification gives a geometrical origin to the Weyl connection and assignment of units to dimensioned quantities. \\

\noindent Using (\ref{meteq}) and (\ref{nonmet}) the transformation rules of co-frame and connection 1-forms read 
\begin{equation}
e^a \mapsto \exp(\sigma) e^a, \quad \Lambda^a_{\ b} \mapsto \Lambda^a_{\ b} - \eta^a_{\ b} d\sigma, 
\end{equation}
respectively. Then, torsion 2-forms and curvature 2-forms transform as
\begin{equation}
T^a \mapsto \exp(\sigma) T^a, \quad
\overset{(\Lambda)}{R^a_{\ b}} \mapsto \overset{(\Lambda)}{R^a_{\ b}},
\end{equation}
respectively. The Ricci 1-forms and the curvature scalar are obtained by contracting the curvature 2-forms:
\begin{equation}
\overset{(\Lambda)}{Ric_a} = \iota_b \overset{(\Lambda)}{R^b_{\ a}}, \ \  \overset{(\Lambda)}{\mathcal{R}} = \iota^a \overset{(\Lambda)}{Ric_a} = \iota^{ab} \overset{(\Lambda)}{R_{ab}}.
\end{equation}
Moreover, the Einstein 2-forms of our non-Riemannian connection are defined by 
\begin{equation}
\overset{(\Lambda)}{G_a}= \overset{(\Lambda)}{G_{ab}} *e^b = -\frac{1}{2} \overset{(\Lambda)}{R_{bc}} \wedge * e^{abc} . \label{efor}
\end{equation}

\section{Scale Covariant Generalisations of Gravitational Theories}

\noindent The gravitational theories that will be discussed below are going to be defined using an action principle. The field equations are going to be derived using a first order variational formalism. For the locally Weyl covariant gravitational theories, the action functional will be given by 
\begin{equation}
I[e^a, \Omega^a_{\ b}, Q^a_{\ b}, \alpha, \lambda_a ] = \int_M \mathcal{L}
\end{equation}
where $M$ is a compact region without boundary on some chart on an $n$-dimensional RCW manifold. The independent variables are the co-frame 1-forms $\{e^a\}$, anti-symmetric part of connection 1-forms $\{ \Omega^a_{\ b} \}$, symmetric part of connection 1-forms $\{Q^a_{\ b}\}$, a real scalar field $\alpha$ with Weyl charge 1 (whose inverse may be regarded as the local gravitational coupling strength), and some Lagrange multiplier valued 1-forms $\{\lambda_a\}$. \\

\noindent To obtain a scale covariant generalisation, we introduce two ingredients to the original theories. First element is the dilaton field $\alpha$. It has the dimensions of inverse length and is used to write scale invariant terms for Lagrangian. The second one is the Weyl connection 1-form $Q$. It is an independent variable through the symmetric part of connection 1-forms. The consistency of the generalisation is checked by following the subsequent diagram:

\begin{center}
\begin{tikzcd}[column sep=1in, row sep=1in]
\mathcal{L}_T \arrow[r, "\text{introduce }\alpha\ \& \ Q"] \arrow[d, "\text{variation}"'] & \mathcal{L}_{SCT} \arrow[d, "\text{variation}"] \\
\dot{\mathcal{L}}_T  \arrow[r, leftarrow, "\alpha=1 \ \&\ Q=0"] &  \dot{\mathcal{L}}_{SCT} 
\end{tikzcd}
\end{center}
We introduce scale invariant terms to the Lagrangian $\mathcal{L}_T$ of the original theory using the dilaton field $\alpha$ and Weyl connection 1-form $Q$. Then vary the scale invariant Lagrangian $\mathcal{L}_{SCT}$ and obtain the scale covariant variational field equations $\dot{\mathcal{L}}_{SCT}$. If these field equations agree with the field equations of original theory $\dot{\mathcal{L}}_T$ for a fixed scale $\alpha=1$, and vanishing non-metricity $Q=0$, the above diagram commutes and we say that the generalisation is consistent. \\

\noindent A consistent generalisation means that the scale covariant theory contains the original theory in its vacuum configuration for the Weyl sector. The vacuum configuration means the Weyl connection 1-form has a vanishing field strength, i.e. it is flat. In this case, any solution of the original theory defines an equivalence class of solutions for the scale covariant theory. In this class, two solutions are related to each other by a pure gauge transformation. \\

\noindent We will now discuss scale covariant generalisations of GR and TMG. We first give the original theories and then discuss their generalisations. 

\section{General Relativity in Three Dimensions}

\noindent We will start with the formulation of GR in 3 dimensions. In our formulation, we use differential forms and first order variational formalism. Our independent variables are the co-frame 1-forms $\{e^a\}$, connection 1-forms $\{ \Omega^a_{\ b} \}$, and Lagrange multiplier valued 1-forms $\{\lambda_a\}$. The Lagrangian density 3-form is given by:
\begin{equation}
\mathcal{L}_{GR}= \frac{1}{K} \overset{(\Omega)}{R^a_{\ b}} \wedge *e_a^{\ b} + \Lambda *1 + \lambda_a \wedge T^a 
\end{equation}
where $K$ is the three dimensional gravitational constant, $\Lambda$ is the cosmological constant. In this theory, we work with the totally anti-symmetic metric compatible connection 1-forms\footnote{This choice of anti-symmetric connections can also be implemented by using Lagrange multipliers, however, it does not make any difference.} 
\begin{equation}
\Omega^a_{\ b}= \omega^a_{\ b}+K^a_{\ b}. \label{con}
\end{equation}
The total variational derivative of $\mathcal{L}_{GR}$ with respect to three independent variables is found to be:
\begin{align}
\dot{\mathcal{L}_{GR}} &= {\dot{e}}^a \wedge \bigg\{ \frac{1}{K} \overset{(\Omega)}{R^b_{\ c}} \epsilon_{ab}^{\ \ c} +\Lambda *e_a + \overset{(\Omega)}{D}\lambda_a \bigg\} + \dot{\lambda_a}\wedge T^a  \nonumber\\
&+\dot{\Omega^a_{\ b}} \wedge \bigg\{\epsilon^a_{\ bc}T^c+\frac{1}{2}(e^b \wedge \lambda_a - e_a \wedge \lambda^b) \bigg\}.
\end{align}
Above, a dot over a field variable denotes the variation of that variable. As a result of the Lagrange constraint equation, torsion vanishes, i.e. $T^a=0$. Putting this in connection variation equation we find the Lagrange multiplier 1-forms as $\lambda_a=0$ identically. Finally, the Einstein field equations are given as:
\begin{equation}
\frac{1}{K}\overset{(\omega)}{R^{ab}}+\frac{\Lambda}{2}e^{ab}=0. \label{gr1}
\end{equation}

\noindent Now we move on to discuss the scale covariant generalisation of GR. In order to promote GR into a locally scale covariant theory, we introduce a real scalar field $\alpha$ with Weyl charge 1 (whose inverse may be regarded as the local gravitational coupling strength). We also take into account the most general connection 1-forms
\begin{equation}
\Lambda^a_{\ b}=\omega^a_{\ b}+K^a_{\ b}+q^a_{\ b}+Q^a_{\ b}
\end{equation}
where the symmetric part $Q^a_{\ b}$ is identified with the Weyl connection 1-form $Q$ through (\ref{nonmet}). Both $\alpha$ and $Q$ are independent variables with respect to which Lagrangian 3-form will be varied. For the variation of Weyl connection we note that 
\begin{equation}
\dot{Q}=-\frac{1}{3}\eta^b_{\ a} \dot{\Lambda^a_{\ b}}.
\end{equation}
We consider the following Weyl invariant Lagrangian density:
\begin{align}
\mathcal{L}_{WGR}&=\alpha \overset{(\Lambda)}{R^a_{\ b}} \wedge *e_a^{\ b} + \alpha^3\Lambda*1 +\alpha\lambda_a\wedge T^a   \nonumber\\
&-\frac{\gamma}{2\alpha} \mathcal{D} \alpha \wedge * \mathcal{D} \alpha-\frac{\gamma^{\prime}}{2\alpha}dQ \wedge *dQ
\end{align} 
Above, we also added the kinetic terms of the fields $\alpha$ and Weyl connection 1-form $Q$ to promote them to dynamical fields. In addition $\gamma$ and $\gamma'$ are dimensionless coupling constants due to scale invariance. The variational field equations of this theory are given by: 
\begin{align}
&\dot{\mathcal{L}_{WGR}} = {\dot{e}}^a \wedge \bigg\{ \alpha \overset{(\Omega)}{R^b_{\ c}} \epsilon_{ab}^{\ \ c} +\alpha^3\Lambda *e_a + \overset{(\Omega)}{D}(\alpha\lambda_a)+\alpha Q \wedge \lambda_a +\frac{\gamma}{2\alpha}\tau_a[\mathcal{D}\alpha] \nonumber\\
&\quad+\frac{\gamma'}{2\alpha}\hat{\tau_a}[dQ]\bigg\} + {\dot{\Omega}}^a_{\ b} \wedge \bigg\{ \overset{(\Omega)}{D}\bigg( \alpha *e_a^{\ b}\bigg) +\alpha e^b \wedge \lambda_a\bigg\} \nonumber\\
&\quad +\dot{Q} \wedge \bigg\{\alpha \lambda_a \wedge e^a -\gamma*\mathcal{D}\alpha-\gamma' d \bigg(\frac{1}{\alpha}*dQ\bigg)\bigg\}+\dot{\lambda_a} \wedge \big(\alpha T^a \big) \nonumber\\
&\quad+\dot{\alpha}\bigg\{\overset{(\Omega)}{R^a_{\ b}} \wedge *e_a^{\ b} +3\alpha^2 \Lambda*1 +\lambda_a \wedge T^a +\frac{\gamma}{2\alpha^2}\mathcal{D}\alpha \wedge*\mathcal{D}\alpha\nonumber\\
&\quad+\gamma\mathcal{D}\bigg(\frac{1}{\alpha}*\mathcal{D}\alpha\bigg)+\frac{\gamma'}{2\alpha^2}dQ\wedge *dQ \bigg\} 
\end{align}
where the shorthand expressions
\begin{align}
\tau_a[\mathcal{D}\alpha]&=(\iota_a\mathcal{D}\alpha)*\mathcal{D}\alpha+\mathcal{D}\alpha \wedge \iota_a *\mathcal{D}\alpha, \\
\hat{\tau_a}[dQ]&=\iota_adQ \wedge *dQ-dQ (\iota_a *dQ),
\end{align}
are the scale covariant stress-energy forms of the dilaton field $\alpha$ and the Weyl vector boson field $Q$, respectively. \\

\noindent We first note from Lagrange constraint equation, since $\alpha \neq 0$, that the torsion 2-forms identically vanish, i.e. $T^a=0$. The field equations will be solved under this constraint. The scalar field equation can be replaced by a simpler expression. In order to demonstrate this, we compare the trace\footnote{Trace of co-frame equation follows from left exterior multiplication with $e^a$.} of the co-frame equation to the $\alpha$ variation equation and obtain:
\begin{equation}
d(\alpha e_a \wedge \lambda^a+\gamma*\mathcal{D}\alpha)=0.
\end{equation} 
Next, we will solve the anti-symmetric connection equation. To do this, we lower an index using 
\begin{equation}
\overset{(\Omega)}{D}(\alpha*e_a^{\ b})=\overset{(\Lambda)}{D}(\alpha*e_a^{\ b})=\mathcal{D}\alpha \wedge *e_a^{\ b}. \label{conlow}
\end{equation}
Then we write the anti-symmetric part of the connection equation as
\begin{equation}
\frac{\alpha}{2}(e_a\wedge \lambda_b - e_b \wedge \lambda_a)=\Sigma_{ab}, \label{ancon}
\end{equation}
where the shorthand expression read:
\begin{equation}
\Sigma_{ab}=\mathcal{D}\alpha \wedge *e_{ab}.
\end{equation}
Equation (\ref{ancon}) is a system of nine algebraic equations for nine unknown variables $\{\lambda_a\}$. The unique solution for the Lagrange multiplier 1-forms is given by:
\begin{align}
\lambda_a &=  \frac{2}{\alpha} \iota^m \Sigma_{ma} - \frac{1}{2\alpha} (\iota^{nm}\Sigma_{mn})e_a \nonumber\\
&= \frac{2}{\alpha}\iota_a *\mathcal{D}\alpha.
\end{align}
Finally, after putting the solution for Lagrange multiplier 1-forms, the variational field equations for the scale covariant general relativity theory read:
\begin{equation}
\alpha \overset{(\Omega)}{R^b_{\ c}} \epsilon_{ab}^{\ \ c} +\alpha^3\Lambda *e_a+2\overset{(\Omega)}{D}(\iota_a*\mathcal{D}\alpha)+2Q\wedge (\iota_a*\mathcal{D}\alpha)+\frac{\gamma}{2\alpha}\tau_a[\mathcal{D}\alpha]+\frac{\gamma'}{2\alpha}\hat{\tau_a}[dQ]= 0,\label{wgr1}
\end{equation}
\begin{equation}
(\gamma+4)*\mathcal{D}\alpha+\gamma'd\bigg(\frac{1}{\alpha}*dQ\bigg)=0, \label{wgr2}
\end{equation}
\begin{equation}
(\gamma+4)d*\mathcal{D}\alpha=0. \label{wgr3}
\end{equation}
We note that, the dilaton field equation (\ref{wgr3}) is given by the Weyl vector field equation (\ref{wgr2}) by taking an exterior derivative. To check the consistency of this generalisation, we make the choice 
\begin{equation}
\mathcal{D}\alpha=0 \ \     \Leftrightarrow  \ \ Q=-\frac{d\alpha}{\alpha}.
\end{equation}
As a result $dQ=0$ and we choose a vacuum class of solutions for the Weyl sector. For this choice, Lagrange multiplier 1-forms vanish identically, i.e. $\lambda_a=0$. and equations (\ref{wgr2}) and (\ref{wgr3}) are trivially satisfied. We are only left with the Einstein field equation:
\begin{equation}
\alpha \overset{(\Omega)}{R^b_{\ c}} \epsilon_{ab}^{\ \ c} +\alpha^3\Lambda *e_a= 0.
\end{equation}
We still have a residual gauge freedom for the dilaton field $\alpha$. We fix this residual gauge freedom by setting $\alpha=1$. For this choice $Q=0$ and the connection 1-forms become the unique Levi-Civita connection 1-forms. Furthermore, this choice amounts to choosing a global units frame in which (\ref{wgr1}) reduces to the Einstein field equation (\ref{gr1}) of the original theory.\footnote{In this unit system, $K=1$.} Therefore this way of Weyl gauging the theory of GR is a consistent generalisation. We now move on to formulate TMG theory and its scale covariant version.

\section{Topologically Massive Gravity}

\noindent We now start with the formulation of TMG using differential forms and first order variational formalism. Our independent variables are the co-frame 1-forms $\{e^a\}$, connection 1-forms $\{ \Omega^a_{\ b} \}$, and Lagrange multiplier valued 1-forms $\{\lambda_a\}$. The Lagrangian density 3-form is given by:
\begin{equation}
\mathcal{L}_{TMG}= \frac{1}{\mu}(\Omega^a_{\ b} \wedge d\Omega^b_{\ a} + \frac{2}{3} \Omega^a_{\ b} \wedge \Omega^b_{\ c} \wedge \Omega^c_{\ a}) + \frac{1}{K} \overset{(\Omega)}{R^a_{\ b}} \wedge *e_a^{\ b} + \Lambda *1 + \lambda_a \wedge T^a 
\end{equation}
where $K$ is the three dimensional gravitational constant, $\Lambda$ is the cosmological constant. In this theory, we work with the totally anti-symmetic metric compatible connection 1-forms:
\begin{equation}
\Omega^a_{\ b}= \omega^a_{\ b}+K^a_{\ b}. \label{con}
\end{equation}
The total variational derivative of $\mathcal{L}_{TMG}$ with respect to three independent variables is found to be:
\begin{align}
\dot{\mathcal{L}_{TMG}} &= {\dot{e}}^a \wedge \bigg\{ \frac{1}{K} \overset{(\Omega)}{R^b_{\ c}} \epsilon_{ab}^{\ \ c} +\Lambda *e_a + \overset{(\Omega)}{D}\lambda_a \bigg\} + \dot{\lambda_a}\wedge T^a  \nonumber\\
&+\dot{\Omega^a_{\ b}} \wedge \bigg\{\frac{2}{\mu}\overset{(\Omega)}{R^b_{\ a}}+\epsilon^a_{\ bc}T^c+\frac{1}{2}(e^b \wedge \lambda_a - e_a \wedge \lambda^b) \bigg\}.
\end{align}
As a result of the Lagrange constraint equation, torsion vanishes, i.e. $T^a=0$ and we are working with the unique Levi-Civita connection 1-forms $\{\omega^a_{\ b}\}$. Then the Lagrange multiplier 1-forms are solved from the connection variation equation. Lagrange multiplier 1-forms are given in terms of Schouten 1-forms $\overset{(\omega)}{Y_a}$ of the Levi-Civita connection:
\begin{equation}
\lambda_a=-\frac{4}{\mu}\bigg(\overset{(\omega)}{Ric_a}-\frac{1}{4}\overset{(\omega)}{\mathcal{R}}e_a\bigg)=:-\frac{4}{\mu}\overset{(\omega)}{Y_a} \label{lm}
\end{equation}
Putting this expression in the co-frame equation, we obtain the Einstein field equations of the original TMG theory as:
\begin{equation}
-\frac{2}{K} \overset{(\omega)}{G_a} +\Lambda *e_a -\frac{4}{\mu}\overset{(\omega)}{C_a}= 0.
\end{equation}
$\overset{(\omega)}{C_a}:=\overset{(\omega)}{D}\overset{(\omega)}{Y_a}$ are the Cotton 2-forms which involve the third order derivatives of the metric.\\

\noindent To promote this theory into a locally scale covariant one, we again introduce the dilaton field $\alpha$ and consider the most general connection 1-forms:
\begin{equation}
\Lambda^a_{\ b}=\omega^a_{\ b}+K^a_{\ b}+q^a_{\ b}+Q^a_{\ b}
\end{equation}
Consequently, the variation of the symmetric part of connection 1-forms $\{Q^a_{\ b}\}$ are related to variation of the Weyl vector 1-form as:
\begin{equation}
\dot{Q^a_{\ b}}=-\eta^a_{\ b}\dot{Q}.
\end{equation}
We analyze the following Weyl invariant Lagrangian density:
\begin{align}
\mathcal{L}_{WTMG}&=\frac{1}{\mu}(\Lambda^a_{\ b} \wedge d\Lambda^b_{\ a} + \frac{2}{3} \Lambda^a_{\ b} \wedge \Lambda^b_{\ c} \wedge \Lambda^c_{\ a}) +\frac{1}{\mu'}Q \wedge dQ +\alpha \overset{(\Lambda)}{R^a_{\ b}} \wedge *e_a^{\ b}    \nonumber\\
&+ \alpha^3\Lambda*1 +\alpha\lambda_a\wedge T^a-\frac{\gamma}{2\alpha} \mathcal{D} \alpha \wedge * \mathcal{D} \alpha -\frac{\gamma^{\prime}}{2\alpha}dQ \wedge *dQ
\end{align}
Above, we added the abelian Chern-Simons term for the Weyl connection 1-form $Q$ in addition to the kinetic terms of the $\alpha$ and $Q$ fields. The total variational derivative of the Lagragian is found to be:
\begin{align}
&\dot{\mathcal{L}_{WTMG}}= {\dot{e}}^a \wedge \bigg\{ \alpha \overset{(\Omega)}{R^b_{\ c}} \epsilon_{ab}^{\ \ c} +\alpha^3\Lambda *e_a + \overset{(\Omega)}{D}(\alpha\lambda_a)+\alpha Q \wedge\lambda_a +\frac{\gamma}{2\alpha}\tau_a[\mathcal{D}\alpha]\nonumber\\
&\quad +\frac{\gamma'}{2\alpha}\hat{\tau_a}[dQ]\bigg\}+ {\dot{\Omega}}^a_{\ b} \wedge \bigg\{ \frac{2}{\mu}\overset{(\Omega)}{R^b_{\ a}} + \overset{(\Omega)}{D}\bigg( \alpha *e_a^{\ b}\bigg) +\alpha e^b \wedge \lambda_a \bigg\}\nonumber\\
&\quad+\dot{Q} \wedge \bigg\{\bigg(\frac{6}{\mu}+\frac{2}{\mu'}\bigg)dQ + \alpha \lambda_a \wedge e^a  - \gamma*\mathcal{D}\alpha  -\gamma'd\bigg(\frac{1}{\alpha}dQ\bigg)\bigg\} \nonumber\\
&\quad+\dot{\alpha}\bigg\{\overset{(\Omega)}{R^a_{\ b}} \wedge *e_a^{\ b} +3\alpha^2 \Lambda*1 +\lambda_a \wedge T^a +\frac{\gamma}{2\alpha^2}\mathcal{D}\alpha \wedge*\mathcal{D}\alpha\nonumber\\
&\quad+\gamma\mathcal{D}\bigg(\frac{1}{\alpha}*\mathcal{D}\alpha\bigg) +\frac{\gamma'}{2\alpha^2}dQ\wedge *dQ \bigg\} +\dot{\lambda_a} \wedge \big(\alpha T^a \big). 
\end{align}
\noindent First, from the Lagrange constraint equation, torsion 2-forms identically vanish. Then, by comparing the trace of the co-frame equation to the dilaton field equation, we obtain:
\begin{equation}
d(\alpha e_a \wedge \lambda^a+\gamma*\mathcal{D}\alpha)=0.
\end{equation} 
Next, we solve the anti-symmetric part of connection equation for Lagrange multiplier 1-forms. To lower an index in this equation, we use (\ref{conlow}) and write the anti-symmetric part of the connection equation as
\begin{equation}
\frac{\alpha}{2}(e_a\wedge \lambda_b - e_b \wedge \lambda_a)=-\frac{2}{\mu}\overset{(\Omega)}{R_{ab}}+\mathcal{D}\alpha \wedge *e_{ab}. \label{ancon2}
\end{equation}
From above, the unique solution for the Lagrange multiplier 1-forms is found to be:
\begin{equation}
\lambda_a = -\frac{4}{\mu\alpha}\overset{(\Omega)}{Y_a}+ \frac{2}{\alpha} \iota_a*\mathcal{D}\alpha.
\end{equation}
The Schouten 1-forms of the anti-symmetric part of connection 1-forms $\{\Omega^a_{\ b}=\omega^a_{\ b}+q^a_{\ b}\}$ are given by:
\begin{equation}
\overset{(\Omega)}{Y_a}= \iota^b\overset{(\Omega)}{R_{ba}}-\frac{1}{4}(\iota^{cb}\overset{(\Omega)}{R_{bc}})e_a.
\end{equation}
Finally,  the variational field equations of the scale covariant TMG theory read:
\begin{align}
-2\alpha \overset{(\Omega)}{G_a} &+\alpha^3\Lambda *e_a -\frac{4}{\mu}\overset{(\Omega)}{D}\overset{(\Omega)}{Y_a}-\frac{4}{\mu}Q\wedge\overset{(\Omega)}{Y_a}+2\overset{(\Omega)}{D}(\iota_a*\mathcal{D}\alpha)\nonumber\\
&+2Q \wedge (\iota_a*\mathcal{D}\alpha)+ \frac{\gamma}{2\alpha}\tau_a[\mathcal{D}\alpha] +\frac{\gamma'}{2\alpha}\hat{\tau_a}[dQ]= 0, \label{wtmg1}
\end{align}
\begin{equation}
(\gamma+4)*\mathcal{D}\alpha=\bigg(\frac{2}{\mu}+\frac{2}{\mu'}\bigg)dQ-\gamma'd\bigg(\frac{1}{\alpha}*dQ\bigg),\label{wtmg2}
\end{equation}
\begin{equation}
(\gamma+4)d*\mathcal{D}\alpha=0.\label{wtmg3}
\end{equation}
To check the consistency of this generalisation, we choose a vacuum configuration for the Weyl sector:
\begin{equation}
\mathcal{D}\alpha=0 \ \     \Leftrightarrow  \ \ Q=-\frac{d\alpha}{\alpha}.
\end{equation}
For this choice, equations (\ref{wtmg2}) and (\ref{wtmg3}) are trivially satisfied. We are only left with the equation:
\begin{equation}
-2\alpha \overset{(\Omega)}{G_a} +\alpha^3\Lambda *e_a -\frac{4}{\mu}\overset{(\Omega)}{C_a}+\frac{4}{\mu\alpha}d\alpha\wedge\overset{(\Omega)}{Y_a}= 0. \label{wtmg4} 
\end{equation}
Fixing the residual gauge freedom by setting $\alpha=1$ annihilates the Weyl connection 1-form, i.e. $Q=0$ and we are left with the unique Levi-Civita connection 1-forms $\{\omega^a_{\ b}\}$. Consequently the last term in equation (\ref{wtmg4}) vanishes, the third term becomes the Cotton 2-form of Levi-Civita connection and the field equations consistently reduce to the field equation of original TMG theory:
\begin{equation}
-2\overset{(\omega)}{G_a} +\Lambda *e_a -\frac{4}{\mu}\overset{(\omega)}{C_a}=0.
\end{equation} 

\section{Conclusion}
In our work, we considered the locally scale covariant generalisations of GR and TMG theories. To achieve scale covariance, we adopted the natural framework provided by a RCW space-time with a non-metricity tensor that has vanishing trace-free part. This framework is natural in the sense that, the purely geometrical non-metricity tensor provides us with a connection $\nabla$ which stays inert under local scale transformations. We then derived the Einstein field equations using a first order variational formalism and discussed the consistency of these generalisations. To this aim, we picked a vacuum configuration, where the stress 2-form of the Weyl connection vanishes, and showed the originial theory lies in this vacuum sector. Therefore, we ensured this method of Weyl gauging provides a consistent generalisation. A scale covariant version of TMG is studied in [15] using a similar technique. However, contrary to our approach, the authors interpret the Weyl vector boson as a potential for the gauge group $U(1)$ of electrodynamics. Since Weyl connection 1-form is a purely real we do not do that. Also, when interpreted as a linear connection, it has a geometrical meaning over space-time. For a  $U(1)$ connection this geometrical meaning is rather obscure because it is a complex valued 1-form. \\

\noindent For future studies, one can couple a Higgs like potential for the dilaton field and study a symmetry breaking scheme for the scale group. This way, one can introduce dimensionful quantities to the theory. Alternatively, the same method can be applied to other three dimensional massive gravity theories with an action formulation such as New Massive Gravity [16, 17], Minimal Massive Gravity [18, 19] or New Improved Massive Gravity [20, 21] theories. Three dimensional applications are also important because of novel topological and geometrical properties of three dimensional Einstein-Weyl spaces [22, 23]. We plan to discuss these in upcoming papers. Another relevant direction that may be considered is to investigate solutions of the scale covariant TMG theory that might generalize the BTZ blackhole solution. We expect our results will prove useful in these contexts.

\end{document}